
\documentstyle{amsppt}
\magnification 1200
\NoBlackBoxes
\document

\centerline {\bf Difference equations with elliptic coefficients}
\centerline {\bf and quantum affine algebras}
\vskip .15 in
\centerline{\bf Pavel I. Etingof}
\vskip .1in
\centerline{Department of Mathematics}
\centerline{Yale University}
\centerline{New Haven, CT 06520}
\centerline{e-mail: etingof\@ math.yale.edu}
\vskip .1in
\centerline{December 7, 1993}
\vskip .1in

\heading
{\bf Introduction}
\endheading

The purpose of this paper is to introduce and study
a $q$-analogue of the holonomic system of differential equations
associated to the Belavin's classical $r$-matrix (elliptic $r$-matrix
equations), or, equivalently, to define an elliptic deformation of the
quantum Knizhnik-Zamolodchikov equations invented by
Frenkel and Reshetikhin \cite{FR}.
In \cite{E}, it was shown that solutions of the elliptic
$r$-matrix equations admit a representation as traces of products
of intertwining operators between certain modules over the Lie
algebra $\hat\frak{sl}_N$. In this paper, we generalize this
construction to the case of the quantum algebra
$U_q(\hat\frak{sl}_N)$.

The main object of study in the paper is a family
of meromorphic matrix functions of $n$ complex variables $z_1,...,z_n$
and three additional parameters $p,q,s$ --
(modified) traces of products of intertwiners between
$U_q(\hat\frak{sl}_N)$-modules. They are a new class of
transcendental functions which can be degenerated into many
interesting special functions -- hypergeometric and
$q$-hypergeometric functions, elliptic and modular functions,
transcendental functions of an elliptic curve, vector-valued modular
forms, solutions of the Bethe ansatz equations etc.

The main result of the paper (Theorem 4.3) states that these functions
satisfy two holonomic systems of difference equations, (4.18) and (4.19) --
the first one has shift parameter $p$ and elliptic modulus $s$,
and the second one has shift parameter $s$ and elliptic modulus $p$.

In Section 1, we recall some properties quantum affine algebras
to be used in subsequent sections. Our exposition follows
\cite{FR}. At the end of the section, we introduce the quantum
analogue of the Sugawara construction for the element $L_0$
of the Virasoro algebra. A similar construction is described in
\cite{ITIJMN}.

In Section 2, we define intertwiners $\Phi(z): M_{\lambda,k}\to
M_{\nu,k}\otimes V_z$, where $M_{\lambda,k}$ is a Verma
module and $V_z$ is an evaluation representation of
the quantum affine algebra ($z\in \Bbb C^*$). Following \cite{FR},
we classify such intertwiners and prove a difference
equation for them (due to Frenkel and Reshetikhin).
Our proof uses the quantum Sugawara construction
and is less technical than the original proof in \cite{FR},
although it relies on essentially the same ideas.
At the end of the Section, we reproduce the proof of the quantum
Knizhnik-Zamolodchikov equations for correlation functions
given in \cite{FR}.

In Section 3, we introduce the outer automorphism $\beta$ of
$U_q(\hat\frak{sl}_N)$ corresponding to the rotation of the Dynkin diagram
(which is an $N$-gon), and define the corresponding operator $B$
acting on Verma modules. We define a class of meromorphic functions of
$n$ complex variables -- traces
of products of
intertwiners twisted by the operator $B$.
We show that these traces satisfy a holonomic system of difference equations
which is a $q$-deformation of the elliptic $r$-matrix system.
This system is equivalent to a system of difference equations with elliptic
coefficients, and the shift parameter $p$ in this system is
independent of the (multiplicative) period $s$ of the elliptic
function occuring in its coefficients.
Systems of difference equations of this type
were considered in \cite{JMN}, where they arose as systems satisfied by
correlation functions of the 8-vertex model in statistical mechanics.
It would be very interesting and useful to clarify the connection between
the difference equations of \cite{JMN} and the difference equations
of this paper, but at the moment it is not clear how to do it.

In Section 4, we define the fundamental trace
-- a matrix-valued function whose columns
are traces of linearly independent intertwiners.
We derive connection relations for the fundamental trace
which are the $q$-analogue of the monodromy for the elliptic $r$-matrix
equations computed in \cite{E}, and the elliptic analogue of the
connection matrices for the quantum KZ equations (see \cite{FR}).
These connection relations turn out to be equivalent to another
holonomic system of
difference equations with elliptic coefficients, in which the
shift parameter and the elliptic modulus interchange.
Thus, we obtain a pair of holonomic systems of difference equations
with elliptic coefficients such that each of them
is the monodromy (= the set connection relations) for the other system.
We call such a pair a double difference system.

In Section 5 we describe some general properties of double difference
systems with $N$-dimensional matrix coefficients.
We start with conventional holonomic systems and show that
a nondegenerate system with rational coefficients
is always solvable in meromorphic functions, whereas
for systems with elliptic coefficients in more than one variable this
is not true. However, if a system with elliptic coefficients is
solvable, and one fixes a matrix solution of it,
 then one can construct a dual system satisfied by this solution,
also with elliptic
coefficients, which is a system of connection relations
for the previous system, and in which the shift parameter and the
elliptic modulus interchange. In this way we obtain consistent
double difference systems with elliptic coefficients.
We give a complete classification of such systems in one variable and
one dimension, and compute their solutions.
This result can be extended to many variables, but
in more than one dimensions the classification is hardly possible.
Even the problem of finding a dual system to a given difference system
with elliptic coefficients (so that they make a consistent double
difference system together) is a difficult trancendental
problem (which is no surprise since it is a generalization of the
notoriously difficult problems of finding connections matrices
for difference systems with rational coefficients, and of computing
monodromy of differential equations). In fact, the example
given in this paper appears to be the first nontrivial explicit example
of a consistent double difference system with elliptic coefficients.

In the appendix, we briefly discuss some limiting cases of the
difference equations deduced in the preceding sections
and their connections to qKZ equations, elliptic KZ equations,
Smirnov's equations, and Bethe ansatz equations.

{\bf Acknowledgements} I would like to thank my advisor Professor
Igor Frenkel for formulating the problem and stimulating my work,
and also I.Cherednik, J.Ding, M.Jimbo, M.Kashiwara, D.Kazhdan,
T.Miwa, Ya. Soibelman, L.Takhtajan,
and A.Varchenko for useful discussions.
\vskip .1in

\heading
{\bf 1. Quantum affine algebras.} \endheading

Let $\frak g$ be a finite dimensional simple Lie algebra over
$\Bbb C$ of rank $r$.
Denote by $<,>$ the standard invariant form on $\frak g$ with respect to which
the longest root has squared length $2$.

Let $\frak h$ denote a Cartan subalgebra of $\frak g$. The form $<,>$
defines a natural identification ${\frak h^*}\to{\frak h}$:
$\lambda\mapsto h_{\lambda}$ for $\lambda\in{\frak h}^*$. We will
use the notation $<,>$ for the inner product in both $\frak h$ and
$\frak h^*$. For the sake of brevity we will
often write $\lambda$ instead of $h_{\lambda}$.

Let $\alpha_i$, $1\le i\le r$ be the simple positive roots of $\frak g$.
Let $\theta$ be the highest root of $\frak g$ -- the positive root
such that $\theta+\alpha_i$ is not a root for any $i$.
Extend the Cartan subalgebra $\frak h$ by adding a new element $c$
orthogonal to $\frak h$. Denote the Cartan subalgebra extended by $c$
by $\hat{\frak h}$: $\hat{\frak h}={\frak h}\oplus \Bbb Cc$.

Introduce the convenient notation $\alpha_0=-\theta$,
$H_i=h_{\alpha_i}$, $1\le i\le r$, $H_0=-h_{\theta}+c$.
Let $a_{ij}=\frac{2<\alpha_i,\alpha_j>}{<\alpha_i,\alpha_i>}$, $0\le
i,j\le r$.

Let $\rho$ be the half sum of positive roots of $\frak g$, and
let $N=1+<\rho,\theta>$ be the dual Coxeter number of $\frak g$.

Let $t$ be a complex number and $q=e^t$. We assume that $|q|<1$.
If $A$ is a number or an operator, by $q^A$ we will always mean
$e^{tA}$.

Let $U_q(\hat{\frak g})$ be the quantum affine
algebra corresponding to $\frak g$ \cite{Dr1,J}. This algebra with unit
is generated
by elements $e_i$, $f_i$, $0\le i\le r$, $q^{\pm h}$, $h\in\hat{\frak
h}$,
satisfying the standard relations
$$
\gather
q^{h_1}q^{h_2}=q^{h_1+h_2},\ h_1, h_2\in \hat{\frak h};\ q^0=1;\\
q^he_iq^{-h}=q^{\alpha_i(h)}e_i,\
q^hf_iq^{-h}=q^{-\alpha_i(h)}f_i,\
, 0\le i\le r,\\
e_if_j-f_je_i=\delta_{ij}\frac{q^{H_i}-q^{-H_i}}{q_i-
q_i^{-1}},\ 0\le i,j\le r,\\
\sum_{n=1}^{1-a_{ij}}(-1)^n\biggl(\matrix 1-a_{ij}\\
n\endmatrix\biggr)_{q_i}e_i^ne_je_i^{1-a_{ij}-n}=0,\\
\sum_{n=1}^{1-a_{ij}}(-1)^n\biggl(\matrix 1-a_{ij}\\
n\endmatrix\biggr)_{q_i}f_i^nf_jf_i^{1-a_{ij}-n}=0,\\
q_i=q^{2/<\alpha_i,\alpha_i>},\ 0\le i,j\le r, i\ne
j,\tag 1.1\endgather
$$
where $\biggl(\matrix m\\
n\endmatrix\biggr)_q$ is the $q$-binomial coefficient: by definition,
$$
\biggl(\matrix m\\
n\endmatrix\biggr)_q=\frac{\prod_{k=m-n+1}^m(q^k-q^{-k})}{\prod_{k=1}^n
(q^k-q^{-k})}.\tag
1.2
$$
The comultiplication in $U_q(\hat{\frak g})$ is defined as
follows:
$$
\gather
\Delta(q^h)=q^h\otimes q^h,\ h\in\hat{\frak h},\\
\Delta(e_i)=e_i\otimes q^{H_i}+1\otimes e_i,\\
\Delta(f_i)=f_i\otimes 1+q^{-H_i}\otimes f_i,\ 0\le i\le
r.\tag 1.3
\endgather
$$
The counit is given by
$$
\epsilon(e_i)=\epsilon(f_i)=0,\ 0\le i\le r,\ \epsilon(q^h)=1,\
h\in\hat{\frak h}.\tag 1.4
$$
The antipode is given by
$$
S(e_i)=-e_iq^{-H_i},\ S(f_i)=-q^{H_i}f_i,\
0\le i\le r,\ S(q^h)=q^{-h},\ h\in\hat{\frak h}.\tag 1.5
$$
Equipped with all these structures, $U_q(\hat{\frak g})$ becomes a
Hopf algebra.

Let us add a new element $D$ to $U_q(\hat{\frak g})$ satisfying
the relations
$$
\gather
[D,e_i]=e_i,\ [D,f_i]=-f_i,\ 0\le i\le r,\ [D,h]=0, h\in\hat{\frak h}\\
\Delta(D)=D\otimes 1+1\otimes D,\ \epsilon(D)=0,\ S(D)=-D .\tag
1.6\endgather
$$

The element $D$ defines a grading on $U_q(\hat{\frak g})$: an element
$a$ is of degree $n$ if $[D,a]=na$. An element is called homogeneous if it is
an eigenvector of the operator of commutation with $D$.

We denote the Cartan subalgebra $\hat{\frak h}$ extended by $D$ by
$\tilde{\frak h}$, and the Hopf algebra $U_q(\hat{\frak g})$ extended by $D$ by
$U_q(\tilde{\frak g})$.

Let us define two classes of modules over $U_q(\tilde{\frak g})$: Verma
modules and evaluation representations (our definition is slightly
different from the standard one; cf. \cite{E}).

The definition of a Verma
module of highest weight $\lambda$ and level $k$ is slightly twisted:
$M_{\lambda,k}$, $(\lambda\in{\frak h^*},k\in\Bbb C)$, is the module
 freely generated by a
vector $v$ satisfying
$$e_iv=0,\ q^hv=q^{<\lambda+k\rho, h>}v,\ h\in{\frak
h},\ q^cv=q^{Nk}v,\ Dv=\frac{<\lambda,\lambda>}{2(k+1)}v.\tag 1.7
$$

Evaluation representations are defined as follows. Let $V$ be a
representation of $U_q(\hat{\frak g})$ which
belongs to category $\Cal O$ when restricted to the
subalgebra $U_q(\frak g)$ generated by $e_i,f_i$ $i\ge 1$, $q^h$,
$h\in{\frak h}$.
 Consider the representation
of $U_q(\tilde{\frak g})$ in the space of $V$-valued Laurent
polynomials $V\otimes\Bbb C[z,z^{-1}]$, defined by
$$
D\circ w(z)=z\frac{dw(z)}{dz};\ a\circ w(z)=z^n\pi_V(a)w(z),\ a\in
U_q(\tilde{\frak g}), \ \text{deg}(a)=n.\tag 1.8
$$
This representation is reducible. Let $V(z)$ denote the
submodule in it spanned by the vectors $v\otimes z^l$, where $v\in V$ runs
over all vectors of such weights $\lambda$ that $<\lambda,\rho>-l$
is divisible by $N$. Then $V(z)$ is called the
Laurent polynomial
 representation of the quantum affine algebra associated to $V$.

If $z_0$ is a nonzero complex number, then the space of all elements of
$V(z)$ vanishing at $z_0$ is a $U_q(\hat{\frak g})$-
submodule of $V(z)$. The corresponding
quotient is denoted by $V_{z_0}$. We will say that $V_{z_0}$ is the
 evaluation representation (at $z_0$) associated to $V$.
In particular, if $z_0=1$ then $V_{z_0}=V$, so $V$ itself is also
an evaluation representation.

The universal quantum $R$-matrix for $U_q(\tilde{\frak g})$ is defined as
follows. Consider the subalgebras $U_q(\frak n^+)$ and $U_q(\frak
n^-)$ in $U_q(\hat{\frak g})$ generated by $\lbrace e_i\rbrace$ and $\lbrace
f_i\rbrace$, respectively. It is known \cite{Dr1} that there exists a
unique pairing between these two algebras, $<,>_t$ such that
$<e_i,f_j>_t=t^{-1}\delta_{ij}$, and
$$
\gather
<ab,c>_t=\sum_j<a,r_i>_t<b,p_i>_t,\ \Delta(c)=\sum_jp_j\otimes r_j,
a,b\in U_q({\frak n}^+),\ c\in U_q({\frak n}^-),\\
<c,ab>_t=\sum_j<p_j,a>_t<r_j,b>_t,\ \Delta(c)=\sum_jp_j\otimes r_j,
a,b\in U_q({\frak n}^-),\ c\in U_q({\frak n}^+)
\tag
1.9
\endgather
$$
Let $\lbrace a_i\rbrace$ be a basis of $U_q(\frak n^+)$ consisting of
homogeneous vectors, and let
$\lbrace a^i\rbrace$ be the basis of $U_q(\frak n^-)$ dual to
 $\lbrace a_i\rbrace$. Then the universal $R$-matrix is given by
$$
\tilde\Cal R=q^{c\otimes d+d\otimes c+\sum_{j=1}^rx_j\otimes
x_j}\sum_ia_i\otimes a^i,\tag 1.10
$$
where $\lbrace x_j\rbrace$ is an orthonormal basis of $\frak h$, and
$d=\frac{1}{N}(D-\rho)$.

The $R$-matrix should be regarded as an infinite expression which
makes sense as an operator in the tensor product
of Verma modules. Such expressions form an algebra -- a completion
of the tensor square of the quantum affine algebra, and the $R$-matrix
is its element.

The $R$-matrix satisfies the following quasitriangularity axioms:
$$
\gather
\tilde\Cal R\Delta(x)=\Delta^{\text{op}}(x)\tilde\Cal R,\\
(\Delta\otimes \text{Id})(\tilde\Cal R)=\tilde\Cal R_{13}\tilde\Cal R_{23},\\
(\text{Id}\otimes\Delta)(\tilde\Cal R)=\tilde\Cal R_{13}\tilde\Cal
R_{12},
\tag 1.11\endgather
$$
where $\Delta^{\text{op}}$ denotes the opposite comultiplication and
$$
\gather
\tilde\Cal R=\sum_ia_i\otimes b_i,\ \tilde\Cal R^{\text{op}}
=\sum_i b_i\otimes a_i,\\
\tilde\Cal R_{12}=\sum_ia_i\otimes b_i\otimes 1=\tilde\Cal R\otimes 1,\\
\tilde\Cal R_{13}=\sum_ia_i\otimes 1\otimes b_i,\\
\tilde\Cal R_{23}=\sum_i1\otimes a_i\otimes b_i=1\otimes \tilde\Cal R.
\tag 1.12\endgather
$$

The matrix $R$ also satisfies the quantum Yang-Baxter equation:
$$
\tilde\Cal R_{12}
\tilde\Cal R_{13}
\tilde\Cal R_{23}=
\tilde\Cal R_{23}
\tilde\Cal R_{13}
\tilde\Cal R_{12}\tag 1.13
$$

We introduce the modified $R$-matrices
$$
\Cal R=q^{-C\otimes D-D\otimes C}\tilde\Cal R,\ \Cal
R(z)=(z^D\otimes 1)\Cal R(z^{-D}\otimes 1), \ C=\frac{c}{N}
\tag 1.14
$$
Note that $\Cal R(z)$ is a power series in $z$ which includes only
nonnegative powers of $z$.

The matrix $\Cal R$ satisfies a modified version of
the quasitriangularity axioms and quantum Yang-Baxter equations:
$$
\gather
\Cal R\Delta(x)=\Delta_C^{\text{op}}(x)\Cal R,\\
(\Delta\otimes \text{Id})(\Cal R(z))=\Cal
R^C_{13}(z)\Cal
R_{23}(z),\\
(\text{Id}\otimes\Delta)(\Cal R(z))=\Cal R^{-C}_{13}(z)\Cal
R_{12}(z),\\
\Cal R_{12}(z_1/z_2)\Cal R^{C}_{13}(z_1/z_3)\Cal R_{23}(z_2/z_3)=
\Cal R_{23}(z_2/z_3)\Cal R^{-C}_{13}(z_1/z_3)\Cal R_{12}(z_1/z_2),
\tag 1.15\endgather
$$
where $\Delta_C(x)=(q^{-CD}\otimes q^{-CD})\Delta(x)(q^{CD}\otimes
q^{CD})$, $\Cal R^{\pm C}_{13}(z)=q^{\pm D\otimes C\otimes 1}\Cal R_{13}(z)
q^{\mp D\otimes C\otimes 1}$.

Unlike $\tilde\Cal R$, the element $\Cal R$ has a remarkable property:
it can be projected to evaluation representations.
Namely, for any $z\in \Bbb C^*$ and an evaluation representation
$V_z$ one can define
the elements
$$
\gather
L^+_V(z)=(\text{Id}\otimes\pi_{V_z})(\Cal
R^{\text{op}})=(\text{Id}\otimes\pi_V)(\Cal R^{\text{op}}(z))\in U_q(\hat{\frak
g})\hat\otimes \text{End}(V),\\
L^-_V(z)=(\text{Id}\otimes\pi_{V_z})(\Cal
R^{-1})=(\text{Id}\otimes\pi_V)(\Cal R^{-1}(z^{-1}))\in U_q(\hat{\frak
g})\hat\otimes \text{End}(V),\tag 1.16
\endgather
$$
where $\hat\otimes$ denotes a completed tensor product.
These elements are called quantum currents.

If we pick a basis in $V$
labeled by a set $I$ then the quantum currents can be viewed as
matrices, $L^{\pm}_{ij}$, $i,j\in I$, whose entries are Laurent
polynomials of $z$ with values in
$U_q(\hat{\frak g})$.

Further, one can define the projection of $\Cal R$ into two
evaluation representations $V_{z_1}^1,V_{z_2}^2$:
$$
R^{V^1V^2}(z_1,z_2)=(\pi_{V^1_{z_1}}\otimes\pi_{V^2_{z_2}})(\Cal
R).\tag 1.17
$$
This projection turns out to be a power series in $z=z_1/z_2$ with
only positive powers of $z$ present. This series converges in a
neghborhood of the origin and therefore defines a holomorphic function with
values in $\text{End}(V^1\otimes V^2)$ in the neighborhood of the
origin, which we will write as $R^{V^1V^2}(z)$ or, when no confusion
is possible, simply as $R(z)$. One can show that this
function extends to a meromorphic function in $\Bbb C$ which is a
product of a scalar trnscendental function $\phi$ and a rational
matrix-valued function $\tilde R$: $R^{V^1V^2}(z)=\phi^{V^1V^2}(z)\tilde
R^{V^1V^2}(z)$. The rational function $\tilde R$ regarded as a function
of $y=\log z$ is a trigonometric solution of the quantum Yang-Baxter
equation. It satisfies the unitarity condition $R(z)R_{21}(z^{-1})=1$.

Let us define a product operation applicable to quantum currents.
Let $a=a_1\otimes a_2\in\Cal U_q(\tilde{\frak g})\otimes
\text{End}(V)$, $b=b_1\otimes b_2\in \Cal U_q(\tilde{\frak g})\otimes
\text{End}(W)$. Define a ``product'' of $a$ and $b$ by
$a*b=a_1b_1\otimes a_2\otimes b_2$. It is important to
distinguish this operation from the usual product.

Let us now write down the commutation relations for currents.

\proclaim{Proposition 1.1}
\cite{FR} The following relations between power series
with values in $\text{End}(M_{\lambda,k}\otimes V^1\otimes V^2)$
hold true:
$$
\gather
(1\otimes R(\frac{z_1}{z_2}))L^{+}_{V^1}(z_1)*L^{+}_{V^2}(z_2)
=S_{V^1V^2}(L^{+}_{V^2}(z_2)*L^{+}_{V^1}(z_1))(1\otimes
R(\frac{z_1}{z_2})),\\
(1\otimes R(\frac{z_1}{z_2}))L^{-}_{V^1}(z_1)*L^{-}_{V^2}(z_2)
=S_{V^1V^2}(L^{-}_{V^2}(z_2)*L^{-}_{V^1}(z_1))(1\otimes
R(\frac{z_1}{z_2})),\\
(1\otimes R(\frac{q^{-k}z_1}{z_2}))L^{+}_{V^1}(z_1)*L^{-}_{V^2}(z_2)
=S_{V^1V^2}(L^{-}_{V^2}(z_2)*L^{+}_{V^1}(z_1))(1\otimes
R(\frac{q^kz_1}{z_2}))
,\tag 1.18\endgather
$$
where $S_{V^1V^2}$ denotes the permutation of the $V^1$ and $V^2$
factors.
\endproclaim

To prove this proposition,
it is enough to apply the maps $\pi_{V^1}\otimes \pi_{V^2}\otimes
\pi_{M_{\lambda,k}}$,
 $\pi_{M_{\lambda,k}}\otimes
\pi_{V^1}
\otimes\pi_{V^2}$, and $\pi_{V^1}\otimes
\pi_{M_{\lambda,k}}\otimes\pi_{V^2}$
 to the quantum Yang-Baxter relation for $\Cal
R$.

Let us now describe a quantum analogue of the Sugawara construction.
Let $m:\ U_q(\tilde{\frak g})\otimes U_q(\tilde{\frak g})\to
U_q(\tilde{\frak g})$ be the multiplication map.
Consider the element
$$
u=m((S\otimes\text{Id})(\tilde\Cal R^{\text{op}})).\tag 1.19
$$
Drinfeld showed that this element satisfies the following relations.

\proclaim{Proposition 1.2}\cite{Dr2}
$$
u^{-1}=\sum S^{-1}(b_i^*)a_i^*, \text{ where } \tilde\Cal R^{-1}=\sum
a_i^*\otimes b_i^*;\tag 1.20
$$
$$
uxu^{-1}=S^2(x),\ x\in U_q(\tilde{\frak g});\tag 1.21
$$
$$
\Delta(u)=(u\otimes u)(\tilde\Cal R^{\text{op}}\tilde\Cal R)^{-1}=
(\tilde\Cal R^{\text{op}}\tilde\Cal R)^{-1}(u\otimes u).\tag
1.22
$$
\endproclaim

Thus we have the following proposition.

\proclaim{Proposition 1.3} In the Verma module
$M_{\lambda,k}$
$$u=q^{2D}.\tag 1.23$$
\endproclaim

\demo{Proof} First of all, we have the equality
$q^{2D}xq^{-2D}=S^2(x)$. This equality can be easily checked: it is
enough to check it for the generators
$x=e_i,f_i,q^h$, since both sides of it are
automorphisms of the quantum affine algebra. Hence, it follows from
the previous proposition that $uq^{-2D}$ commutes with the quantum
affine algebra, so it is a constant. To prove that this constant is 1,
it is enough to check that
$uv=q^{2D}v$, where $v$ is the vacuum vector, which is straightforward.
\enddemo

Let
$$
\hat \Cal R=(q^{-2CD}\otimes 1)\Cal R(q^{2CD}\otimes 1).\tag 1.24
$$

\proclaim{Proposition 1.4}(Quantum Sugawara construction) The following
relation is satisfied in any Verma module:
$$
q^{2(C+1)D}=m((S\otimes\text{Id})(\hat \Cal R^{\text{op}}))\tag 1.25
$$
\endproclaim

\demo{Proof} It follows from the definition of $\hat\Cal R$ that the right
hand side of (1.54) is equal to $q^{2CD}u$. But $u=q^{2D}$, so we get
(1.25).
\enddemo
\vskip .1in

\heading
{\bf 2. Intertwining operators and difference equations}
\endheading

We will be interested in $U_q(\tilde{\frak g})$-intertwining
operators $\Phi(z):M_{\lambda,k}\to M_{\nu,k}\hat\otimes
z^{\Delta}V(z)$, where
$\hat\otimes$ denotes the completed tensor product, and $\Delta$ is a
complex number. It turns out that such operators may be nonzero if and
only if $\Delta$ equals $\frac{<\nu,\nu>-<\lambda,\lambda>}{2(k+1)}$
plus an integer. The shift by an integer is unimportant, so we will
assume that $\Delta=\frac{<\nu,\nu>-<\lambda,\lambda>}{2(k+1)}$.

\proclaim{Proposition 2.1} \cite{FR}
Operators $\Phi$ are in one-to-one correspondence
with vectors in $V$ of weight $\lambda-\nu$.
This correspondence is defined by the action of $\Phi$ at the vacuum level.
\endproclaim

Let $z_0$ be a nonzero complex number. Evaluation of the operator
$\Phi(z)$ at the point $z_0$ yields an operator $\Phi(z_0):M_{\lambda,k}\to
M_{\nu,k}\hat\otimes V_{z_0}$.

Sometimes (when no confusion is possible) we will use
the notation $\Phi(z)$ for
the operator
$\Phi$ evaluated at the point $z\in\Bbb C^*$. This will give us an opportunity
to regard the operator $\Phi(z)$ as an analytic function of $z$. This
analytic function will be multivalued: $\Phi(z)=z^{\Delta}\Phi^0(z)$, where
$\Phi^0$ is a single-valued function on $\Bbb C^*$.

Let $W$ be an evaluation representation of $U_q(\hat{\frak g})$. Then
the intertwining property for $\Phi(z)$
can be written in the form
$$
\gather
(\Phi(z)\otimes \text{Id})L_W^+(w)=R_{WV}(q^{-k}w/z)
L^+_W(w)(\Phi(z)\otimes\text{Id})\\
(\Phi(z)\otimes\text{Id})L_W^-(w)=R_{VW}(z/w)^{-1}
L^-_W(w)(\Phi(z)\otimes\text{Id})
. \tag
2.1\endgather
$$

Relations (2.1) are equalities
of maps $M_{\lambda,k}\hat\otimes W\to M_{\nu,k}\hat\otimes V_z\otimes W$.

To prove these formulas, it is enough to combine the second and third relations
of (1.15) with the intertwining relation $\Phi(z)x=\Delta(x)\Phi(z)$.

{\bf Remark. } Note that relations (1.18) and (2.1) are a priori
satisfied only formally, as equalities between power series.
However, since we know that both sides of these equalities
extend to meromorphic
functions, we can conclude that they are also satisfied
analytically for almost all values of the parameters.

Let us now deduce the difference equation for intertwining operators,
following the method of Frenkel and Reshetikhin.

Let
$$
U=q^{2CD}u.\tag 2.2
$$
Introduce the notation $Q=q^{-2(k+1)D}$, $p=q^{-2(k+1)}$. We assume
that $|p|<1$. The quantum Sugawara construction implies that
$Q=U^{-1}$ in $M_{\lambda,k}$. Also, the operator $Q$ acts in
the Laurent series representation $V(z)$ as follows:
$Qv(z)=v(pz)$.

Since $\Phi$ is an
intertwiner, we have the following relation between Laurent series in $z$:
$$
\Phi(z)Q^{-1}=\Delta(Q^{-1})\Phi(z)=(Q^{-1}\otimes Q^{-1})\Phi(z).\tag 2.3
$$
Using the quantum Sugawara formula (1.25) and the identity
$(1\otimes Q^{-1})\Phi(z)=\Phi(p^{-1}z)$, we obtain
$$
\Phi(p^{-1}z)=(Q\otimes 1)\Phi(z)Q^{-1}=
(U^{-1}\otimes 1)\Phi(z)U,\tag 2.4
$$
where $U$ is defined by (2.2).

Introduce the notation $\Phi\bullet (\sum_ia_i\otimes b_i)=\sum_i(1\otimes
b_i)\Phi a_i$, $\Phi\in\text{Hom}_{\Bbb C}(M_{\lambda,k},
 M_{\nu,k}\hat\otimes z^{\Delta}V(z))$.

\proclaim{Lemma 2.2}
$$
(U^{-1}\otimes 1)\Phi(z)U=(L_V^+(q^kz)^{-1}\Phi(z))\bullet
L_V^-(p^{-1}z).\tag 2.5
$$
\endproclaim

\demo{Proof}
We have $U=uq^{2CD}=\sum_jS(b_j)a_jq^{2CD}$.
Therefore, since $\Phi(z)$ is an intertwiner,
$$
\Phi(z)U=\sum_{j}\Delta(S(b_j))\Phi(z)a_jq^{2CD}=
\sum_{j}(S\otimes S)(\Delta^{\text{op}}(b_j))\Phi(z)a_jq^{2CD}
.\tag
2.6
$$

Following Drinfeld (\cite{Dr2}),
introduce the notation $(X\otimes Y\otimes Z)\circ\Phi=
(S(Y)\otimes S(Z))\Phi X$. This defines a right action
of the tensor cube of the quantum affine algebra on the space
$\text{Hom}_{\Bbb C}(M_{\lambda,k}, M_{\nu,k}\hat\otimes z^{\Delta}V(z))$.

Using this notation, we can write (2.6) as follows:
$$
\Phi(z)U=(\text{Id}\otimes
\Delta^{\text{op}})(\tilde\Cal R)\circ \Phi(z)\cdot q^{2CD}.\tag 2.7
$$

Applying (1.11), we get
$$
\Phi(z)U=(\tilde\Cal
R_{12}
\tilde\Cal R_{13})\circ \Phi(z)\cdot q^{2CD}=\tilde\Cal
R_{13}\circ (
\tilde\Cal R_{12}\circ \Phi(z))\cdot q^{2CD}.\tag 2.8
$$

Let us separately consider the expression
$X=\tilde\Cal
R_{12}\circ\Phi(z)$ which
occurs in (2.8). Using the intertwining property of $\Phi(z)$ and (1.11), we
obtain
$$
X=Y\Phi(z),\ Y=m_{31}\bigl((\Delta\otimes S)(\tilde\Cal
R)\bigr),
\tag 2.9
$$
where $m_{31}(a\otimes b\otimes c)=ca\otimes b$.

Applying (1.11), we find
$$
Y=m_{31}\bigl((\text{Id}\otimes\text{Id}\otimes
S)(\tilde\Cal R_{13}\tilde\Cal R_{23})\bigr)=
(S\otimes \text{Id})(\tilde\Cal R^{\text{op}})\cdot (u\otimes 1)
.\tag 2.10
$$

Thus we have
$$
\gather
(U^{-1}\otimes 1)\Phi(z)U=(U^{-1}\otimes 1)\bigl[\tilde\Cal
R_{13}\circ
(S\otimes \text{Id})(\tilde\Cal R^{\text{op}})(u\otimes 1)
\Phi(z)\bigr]q^{2CD}=\\
(\text{using (1.21)})\\
(q^{-2CD}\otimes 1)\bigl[\tilde\Cal R_{13}\circ(S^{-1}\otimes
\text{Id})(\tilde\Cal R^{\text{op}})\Phi(z)\bigr]q^{2CD}=\\
(q^{-2CD}\otimes 1)\bigl[\tilde\Cal R_{13}\circ(\tilde\Cal
R^{\text{op}})^
{-1}\Phi(z)\bigr]q^{2CD}=\\
(q^{-2CD}\otimes 1)
\bigl[(\tilde\Cal R^{\text{op}})^{-1}\Phi(z)\bullet (\text{Id}\otimes
S)(
\tilde\Cal R_{13})\bigr]q^{2CD}=\\
(q^{-2CD}\otimes 1)
\bigl[(\tilde\Cal R^{\text{op}})^{-1}\Phi(z)\bullet
(S^{-2}\otimes\text{Id})
(\tilde\Cal R_{13}^{-1})\bigr]q^{2CD}=\\
(q^{-2kD}\otimes 1)\cdot
 (\Cal R^{\text{op}})^{-1}(1\otimes q^{-kD})(q^{2kD}\otimes
q^{2kD})\Phi(z)\bullet (\Cal R_{13}^{-1}(q^{-2-2k})
(1\otimes q^{-kD}))
=\\
\Cal R_{32}^{-1}(q^k)\Phi(z)\bullet \Cal R_{13}^{-1}(p)=\\
(L_V^+(q^kz)^{-1}\Phi(z))\bullet
L_V^-(p^{-1}z).\tag 2.11\endgather
$$
\enddemo

The lemma together with equation (2.4) implies
the following difference equation for $\Phi(z)$:

\proclaim{Theorem 2.3}
\cite{FR} The intertwining operator $\Phi(z)$ satisfies the
difference equation
$$
\Phi(pz)=L_V^+(pq^kz)(\Phi(z)\bullet L_V^-(z)^{-1}).\tag 2.12
$$
\endproclaim

Let us now deduce the difference equations for quantum correlation
functions. Let $V^1,...,V^n$ be evaluation representations of
$U_q(\hat{\frak g})$, and let
$\Phi^j(z_j): M_{\lambda_j, k}\to M_{\lambda_{j-1},k}\hat\otimes V^j_{z_j}$
be $U_q(\hat{\frak g})$-intertwining operators. Then the product
$\Phi^1(z_1)\dots\Phi^n(z_n)$ makes sense as an operator
$M_{\lambda_n,k}\to M_{\lambda_0,k}\hat\otimes V^1\otimes\dots\otimes
V^n$ if $|z_1|>>|z_2|>>\dots>>|z_n|$. Consider the matrix element of
this product corresponding to the vacuum vectors in the Verma modules:
$$
\Psi(z_1,...,z_N)=<v_{\lambda_0}^*,\Phi^1(z_1)\dots\Phi^n(z_n)v_{\lambda_n}>,
\tag
2.13
$$
where $v_{\lambda_n}$ is the highest weight vector of $M_{\lambda_n}$
and $v_{\lambda_0}^*$ is the lowest weight vector of $M_{\lambda_0}$.
This function takes values in the tensor product
$V^1\otimes\dots\otimes V^n$. It is called the quantum correlation
function.

We have
$$
\gather
\Psi(z_1,...,pz_j...,z_n)=<v_{\lambda_0}^*,\Phi^1(z_1)\dots\Phi^j(pz_j)\dots
\Phi^n(z_n)v_{\lambda_n}>=\\
<v_{\lambda_0}^*,\Phi^1(z_1)\dots
\Phi^{j-1}(z_{j-1})L^+_{V^j}(pq^kz_j)\Phi^j(z_j)\bullet
L^-_{V^j}(z_j)^{-1}
\Phi^{j+1}(z_{j+1})\dots\Phi^n(z_n)v_{\lambda_n}>.
\tag 2.14\endgather
$$

Let us now drag $L^+$ to the left and $L^-$ to the right, using
commutation relations (2.1). Taking into account the relations
$$
\gather
(S\otimes \text{Id})(L_V^+(z))(v_{\lambda_0}^*\otimes
v)=v_{\lambda_0}^*\otimes q^{\lambda_0}v,\\
L_V^-(z)^{-1}(v_{\lambda_n}\otimes v)=v_{\lambda_n}\otimes
q^{\lambda_n}v,
\tag 2.15\endgather
$$
which follow from the definition of the quantum currents,
 we will get the following result.

\proclaim{Theorem 2.4}\cite{FR}
The quantum correlation functions satisfy the following system of
linear difference equations.
$$
\gather
\Psi(z_1,...,pz_j...,z_n)=R_{jj-1}^{V^jV^{j-1}}(\frac{pz_j}{z_{j-1}})\dots
R_{j1}^{V^jV^{1}}(\frac{pz_j}{z_1})(q^{\lambda_0+\lambda_n})\mid_{V_j}\times
\\
R_{nj}^{V^nV^j}(\frac{z_n}{z_j})^{-1}\dots R_{j+1j}^{V^{j+1}V^j}(\frac
{z_{j+1}}{z_j})^{-1}
\Psi(z_1,...,z_j,...,z_n).\tag 2.16
\endgather
$$
\endproclaim

\heading
{\bf 3. Difference equations for traces of intertwiners.}
\endheading

{}From now on the letter $\frak g$ will denote the Lie algebra
${\frak sl}_N(\Bbb C)$ of traceless $N\times N$ matrices with complex
entries. The dual Coxeter number of this algebra is $N$, and the rank is $N-1$.
The Cartan subalgebra $\frak h$ is the subalgebra of diagonal matrices.

Let $B$ be the $N\times N$ matrix of zeros and ones
corresponding to the cyclic permutation $(12...N)$:
$$
B=\left(\matrix 0 & 0 & \dots & 0 & 1\\ 1 & 0 & \dots & 0 & 0\\
0 & 1 & \dots & 0 & 0\\ \dots &\dots &\dots &\dots &\dots \\
0 & 0 & \dots & 1 & 0\endmatrix\right)
$$

Define an inner automorphism $\beta$ of $\frak h$:
$\beta(a)=BaB^{-1}$, $a\in\frak h$.
This automorphism has order $N$. We can extend it to $\tilde {\frak h}$ by
making it act trivially on $c$ and $D$.

The automorphism $\beta$ can be extended to an
outer automorphism of $U_q(\tilde{\frak g})$ defined
by the relations
$$
\beta(e_i)=e_{i+1},\ \beta(f_i)=f_{i+1},\ \beta(q^h)=q^{\beta(h)},\
h\in\tilde{\frak h},\tag 3.1
$$
where the subscripts are regarded modulo $N$.
This outer automorphism also has order $N$ and can be interpreted as
the rotation of the Dynkin diagram of $\hat {\frak g}$ (which is a regular
$N$-gon)
through the angle $2\pi/N$.

The action of $\beta$ in $U_q(\tilde{\frak g})$ preserves
degree,
hence,
it preserves the polarization. Therefore, it transforms Verma modules into
Verma modules. In other words, we can regard $\beta$ as an operator
$B:M_{\lambda,k}\to M_{\beta(\lambda),k}$, where by convention
$\beta(\lambda)(h)=\lambda(\beta^{-1}(h))$.  This operator intertwines the
usual action of $U_q(\tilde{\frak g})$
and the action twisted by $\beta$:
$\beta(a)Bw=Baw$, $a\in U_q(\tilde{\frak g})$, $w\in M_{\lambda,k}$.

Let $\nu=\beta^{-1}(\lambda)$, and let $\Phi^j(z_j)$ be as above (cf.
section 2).

We assume that the representations $V^j$ are
finite-dimensional and irreducible when restricted to $U_q(\frak g)$.
It is easy to show that for any such representation
$V$ there exists a unique, up to a constant, operator $B:V\to V$
such that $Bav=\beta(a)Bv$ for $a\in U_q(\hat{\frak g})$,
$v\in V$ -- it follows from the fact that $V$ twisted by $\beta$ is
isomorphic to $V$. The operator $B$ gives rise to a well defined automorphism
$\beta$ of $\text{End(V)}$: $\beta(E)=BEB^{-1}$.

Let $s$ be a complex number, $0<|s|<1$. Following the idea of
Bernard \cite{Ber},
Frenkel, Reshetikhin (\cite{FR}, Remark 2.3) (see also \cite{E},\cite{EK}),
introduce the following formal power series in $z_1/z_2$, $z_2/z_3$,...,
$sz_n/z_1$:
$$
F(z_1,...,z_n|s)=\text{Tr}\mid
_{M_{\lambda,k}}(\Phi^1(z_1)\dots\Phi^n(z_n)Bs^{-D}).\tag 3.2
$$
It is not difficult to prove that
this series defines
an analytic function
when $|z_1|>>|z_2|>>\dots>>|z_n|>>|sz_1|$.
This function takes values in $V^1\otimes\dots\otimes V^n$.
{}From now on
it will be the main object of our study.

It turns out that the $n$-point trace $F(z_1,...,z_n|s)$ defined by (3.2)
satisfies a remarkable system of difference equations involving
elliptic solutions of the quantum Yang-Baxter equation for ${\frak sl}_N$.
Let us deduce these equations.

According to (2.14), we have
$$
\gather
F(z_1,...,pz_j...,z_n|s)=
\text{Tr}\mid_{M_{\lambda,k}}(\Phi^1(z_1)\dots\Phi^j(pz_j)\dots
\Phi^n(z_n)Bs^{-D})=\\
\text{Tr}\mid_{M_{\lambda,k}}(\Phi^1(z_1)\dots
\Phi^{j-1}(z_{j-1})L^+_{V^j}(pq^kz_j)\Phi^j(z_j)\bullet
L^-_{V^j}(z_j)^{-1}\Phi^{j+1}(z_{j+1})\dots \Phi^n(z_n)Bs^{-D})
.\tag 3.3\endgather
$$

We would like to deduce a difference equation for $F$.

 First of all, we need to describe some properties of the automorphism
$\beta$. Since $\beta$ is a degree preserving automorphism,
it preserves the universal $R$-matrix:
$$
(\beta\otimes\beta)(\tilde\Cal R)=\tilde\Cal R,\
(\beta\otimes\beta)(\Cal R)=\Cal R.\tag 3.4
$$
Let $\Cal R^M=(\beta^M\otimes 1)(\Cal R)$.
It is obvious that $(\beta^I\otimes \beta^J)(\Cal R)=\Cal R^{I-J}$.

Introduce the following notation:
$$
\gather
L^+_V(z)^M=(\text{Id}\otimes\pi_{V_z})((\Cal
R^M)^{\text{op}})\\
L^-_V(z)^M=(\text{Id}\otimes\pi_{V_z})((\Cal
R^M)^{-1}),\\
R^{V^1V^2}(z)_M=(\pi_{V^1(z)}\otimes\pi_{V_2})(\Cal R^M).
\tag 3.5
\endgather
$$

Let $W$ be a finite dimensional representation of $U_q(\hat{\frak
g})$.
Let $\Phi: M_{\beta(\lambda),k}\to \hat M_{\lambda,k}\otimes W$
be an intertwining operator.
Introduce a new function $\tilde F^{IJ}(x,y|s)$ with values in
$W\otimes \text{End}V\otimes\text{End}V$ defined by
$$
\tilde F^{IJ}(x,y|s)=
\text{Tr}\mid_{M_{\lambda,k}}(L^+_{V}(pq^kx)^I_3*\Phi_1\bullet
(L^-_{V}(y)^J)^{-1}_2Bs^{-D}).
\tag 3.6
$$

Our plan is to show that $\tilde F$ satisfies a system of
difference equations,
by dragging $L^+$
around the circle from
right to left and $L^-$ from left to right using the commutation
relations for quantum currents and the property of trace:
$\text{Tr}(ab)=\text{Tr}(ba)$.
Before we do so, we need a
few identities.

First, we have the following generalization of
the third equation in (1.18):
$$
L_{V^1}^+(z_1)^I_3R_{32}(q^kz_1/z_2)_{J+I}(L_{V^2}^-(z_2)_2^J)^{-1}
=(L_{V^2}^-(z_2)_2^J)^{-1}R_{32}(q^{-k}z_1/z_2)_{J+I}L_{V^1}^+(z_1)^I_3,
\tag 3.7
$$
which implies that
$$
(L_{V^2}^-(z_2)_2^J)^{-1}*L_{V^1}^+(z_1)^I_3=R_{32}^{rl}(q^kz_1/z_2)_{J+I}
R_{32}^{lr}(q^{-k}z_1/z_2)_{J+I}^{-1}L_{V^1}^+(z_1)^I
_3*(L_{V^2}^-(z_2)^J_2)^{-1},\tag
3.8
$$
where $R^{rl}$ implies that the first component of $R$ is applied from
the right and the second one from the left, and $R^{lr}$ implies that
the first component of $R$ is applied from the left, and the second
one from the right.
These relations are equalities between elements of the product
$U_q(\hat{\frak g})\otimes \text{End}V\otimes\text{End}V$.
The notation $L(z)_2$ and $L(z)_3$ implies that the second component
of $L(z)$ operates in the second and third factor of this tensor
product, respectively.
To prove these relations, it is enough
to apply the automorphism $1\otimes\beta^{-I}
\otimes\beta^{-J}$ to the quantum
Yang-Baxter relation for the $R$-matrix, and then project the obtained
relation to the corresponding product of representations of the
quantum affine algebra.

Next, we have the identities
$$
L_V^-(z)^{-1}s^{-D}=s^{-D}L_V^-(s^{-1}z)^{-1},\
s^{-D}L_V^+(z)=L_V^+(sz)s^{-D},\tag 3.9
$$
and
$$
L_V^-(z)^{-1}B=B\beta^{-1}(L_V^-(z)^{-1}),\
BL_V^+(z)=\beta(L_V^+(z))B.\tag 3.10
$$

Finally, we have the following $\beta$-twisted versions of
relations (2.1):
$$
\gather
\Phi_1 L_V^+(x)^I_3=R_{31}(q^{-k}x)_{I}L_V^+(x)^I_3\Phi_1,\\
(L_V^-(y)_2^I)^{-1}R_{12}(y^{-1})_I\Phi_1=\Phi_1(L_V^-(y)_2^I)^{-1},
\tag 3.11\endgather
$$
These relations are equalities of elements of
$\text{Hom}(M_{\lambda,k},M_{\nu,k}\otimes
W)\otimes\text{End}V\otimes\text{End}V$, and the meaning of the
subscripts 1,2,and 3 is as above.

Now we are in a position to compute
$\tilde F$. Combining relations (3.6)-(3.8) with (2.1), we get
$$
\gather
\tilde F^{I+1J}(sx,y|s)=
R^{rl}_{32}(\frac{q^{-2}sx}{y})_{J+I+1}R^{lr}_{32}(\frac{psx}{y})_{J+I+1}^{-1}
R_{31}({psx})_{I+1}\tilde F^{IJ}(x,y|s),\\
\tilde F^{IJ-1}(x,s^{-1}y|s)=
R^{rl}_{32}(\frac{q^{-2}sx}{y})_{J+I-1}R^{lr}_{32}(\frac{psx}{y})_{J+I-1}^{-1}
R_{12}^{lr}(\frac{s}{y})_{J-1}^{-1}
\tilde F^{IJ}(x,y|s).
\tag 3.12\endgather
$$
Let $\Cal G^{IJ}=B^{J}_2B^{-I}_3\tilde F^{IJ}B^{-J}_2B^{I}_{3}$. Then we have
$$
\gather
\Cal G^{I+1J}(sx,y|s)=
R^{rl}_{32}(\frac{q^{-2}sx}{y})R^{lr}_{32}(\frac{psx}{y})^{-1}
R_{31}(psx)B_3^{-1}\Cal G^{IJ}(x,y|s)B_3,\\
\Cal G^{IJ-1}(x,s^{-1}y|s)=
R^{rl}_{32}(\frac{q^{-2}sx}{y})R^{lr}_{32}(\frac{psx}{y})^{-1}
R_{12}^{lr}(\frac{s}{y})^{-1}
B_2^{-1}\Cal G^{IJ}(x,y|s)B_2.
\tag 3.13\endgather
$$

Let
$$
\gather
T_0(x,y)=B_3^lR_{31}(psx)^{-1}R^{lr}_{32}(\frac{psx}{y})R^{rl}_{32}
(\frac{q^{-2}sx}{y})^{-1}
(B_3^r)^{-1},\\
T_1(x,y)=B_2^lR_{12}^{lr}(\frac{s}{y})R^{lr}_{32}(\frac{psx}{y})R^{rl}_{32}
(\frac{q^{-2}sx}{y})^{-1}(B_2^r)^{-1},
\tag 3.14\endgather
$$
and let
$$
P_0(x,y)=T_0(x,y)T_0(sx,y)\dots T_0(s^{N-1}x,y),\
P_1(x,y)=T_1(x,y)T_1(x,s^{-1}y)\dots T_1(x,s^{-N+1}y),\tag 3.15
$$
where $T_0,T_1,P_0,P_1\in\text{End}(W\otimes\text{End}V\otimes\text{End}V)$.
Let $\Cal G^{00}$ be denoted by $\Cal G$. Then
we have the following equation:
$$
\Cal G(x,y|s)=
\prod_{M=0}^{\infty}P_0(x,s^{MN}y)\prod_{M=0}^{\infty}
P_1(0,s^{-MN}y)\Cal
G(0,\infty|s)\tag 3.16
$$

Let us now consider the expression $\Cal G(0,\infty|s)$.
We have
$$
\gather
\Cal G(0,\infty|s)=\\
\text{Tr}\mid_{M_{\lambda,k}}\bigl((q^{\sum
x_i\otimes x_i-C\otimes\rho})_3*\Phi_1\bullet
(q^{-\sum x_i\otimes x_i+C\otimes \rho})_2
Bs^{-D}\bigr)=\\
(q^{k(\rho\otimes
1-1\otimes\rho)})_{23}\text{Tr}\mid_{M_{\lambda,k}}\bigl(
(q^{\sum
x_i\otimes x_i})_3
(q^{-\sum
(x_i\otimes 1+1\otimes x_i)\otimes x_i})_{12}*\Phi_1
Bs^{-D}\bigr)
\tag 3.17\endgather
$$

Apart from that, we have the identity
$$
\gather
\text{Tr}\mid_{M_{\lambda,k}}\bigl(h\Phi_1
Bs^{-D}\bigr)+h\text{Tr}\mid_{M_{\lambda,k}}\bigl(\Phi_1
Bs^{-D}\bigr)=
\\
\text{Tr}\mid_{M_{\lambda,k}}\bigl(\Phi_1
hBs^{-D}\bigr)=\text{Tr}\mid_{M_{\lambda,k}}\bigl(\beta^{-1}(h)\Phi_1
Bs^{-D}\bigr),\ h\in {\frak h},\tag 3.18
\endgather $$
which implies that
$$
\text{Tr}\mid_{M_{\lambda,k}}\bigl(h\Phi_1
Bs^{-D}\bigr)=(\beta^{-1}-1)^{-1}(h)\text{Tr}\mid_{M_{\lambda,k}}\bigl(\Phi_1
Bs^{-D}\bigr).\tag 3.19
$$
Hence, if $\phi$ is any analytic function then
$$
\text{Tr}\mid_{M_{\lambda,k}}\bigl(\phi(h)\Phi_1
Bs^{-D}\bigr)=\phi\bigl((\beta^{-1}-1)^{-1}(h)\bigr)
\text{Tr}\mid_{M_{\lambda,k}}\bigl(\Phi_1
Bs^{-D}\bigr).\tag 3.20
$$
For brevity introduce the notation $\chi=(\beta^{-1}-1)^{-1}$. Then
 expression (3.17) can be rewritten in the form
$$
\gather
\Cal G(0,\infty|s)=\\
q^{1\otimes (k\rho\otimes 1-1\otimes k\rho)+\sum_{i}(\chi(x_i)\otimes 1\otimes
x_i-(1+\chi)(x_i)\otimes x_i\otimes
1)}\biggl[\text{Tr}\mid_{M_{\lambda,k}}
\bigl(\Phi_1
Bs^{-D}\bigr)\otimes \text{Id}\otimes \text{Id}\biggr].\tag 3.21
\endgather
$$

Let
$$
\gather
P(x,y)=\\
\prod_{M=0}^{\infty}P_0(x,s^{MN}y)\prod_{M=0}^{\infty}
P_1(0,s^{-MN}y)q^{1\otimes (k\rho\otimes 1-1\otimes k\rho)+
\sum_{i}(\chi(x_i)\otimes 1\otimes
x_i-(1+\chi)(x_i)\otimes x_i\otimes 1)}\tag 3.22
\endgather
$$

In our situation $W=V_1\otimes V_2\otimes\dots\otimes V_n$, $V=V_i$,
and
$\text{End}V_i$ naturally acts on $W$. Therefore, it makes sense to
consider
the $\text{End}W$-valued function
$$
X_i(z_1,...,z_n|s)=m_{321}(P^{(i)}(z,z)),\tag 3.23
$$
where $P^{(i)}(x,y)\in\text{End}W
\otimes\text{End}V_i\otimes\text{End}V_i$ is defined by
(3.22) with $V=V_i$, and
by definition $m_{321}(a\otimes b\otimes c)=cba$.

Now it remains to observe that equations (2.12), (3.16), (3.21), and
(3.22) imply the following difference equation for the function $F$:

\proclaim{Theorem 3.1} The function $ F(z_1,...,z_n|s)$ satisfies the
difference equations
$$
F(z_1,...,pz_i,...,z_n|s)=X_i(z_1,...,z_n|s)F(z_1,...,z_i,...,z_n|s)\tag
3.24
$$
\endproclaim

Note that the coefficients of these difference equations are
meromorphic functions which we have explicitly represented
as (contracted) infinite products of trigonometric $R$-matrices.
In the next section we will show that they can be expressed
in terms of elliptic functions.

\heading
\bf 4. Monodromy equations
\endheading

The quasi-classical limit (i.e. the limit $q\to 1$) of equations (3.24)
 is the system of elliptic KZ
equations described in \cite{E}. For this system one can define the
notion of monodromy which turns out to be expressed by products
of $R$-matrices. It is a natural question what is the quantum analogue
of monodromy. In this section we will give an answer to this question.
The answer is that the role of monodromy is played by another system
of difference equations with elliptic coefficients which are
products of $R$-matrices depending on spectral parameters.

Let us first describe how to interchange the order of intertwining
operators.

Let $\Phi^{w,\lambda,\nu}(z):M_{\lambda,k}\to M_{\nu,k}\hat\otimes
z^{\Delta}V(z)$ be the intertwining operator such that
\linebreak $<v_{\nu}^*,\Phi^{w,\lambda,\nu}(z)v_{\lambda}>=w$, $w\in
V^{\lambda-\nu}$. Suppose that $z_1,z_2$ are nonzero complex numbers,
and we have a product
$\Phi^{w_1,\lambda_1,\lambda_0}(z_1)\Phi^{w_2,\lambda_2,\lambda_1}(z_2)
:M_{\lambda_2,k}\to M_{\lambda_0,k}\hat\otimes V^1\otimes V^2$ where
$V^1$ and $V^2$ are finite dimensional representations of $U_q(\hat{\frak g})$.
The question is: can this product be expressed in terms of products of
the form $\Phi(z_2)\Phi(z_1)$? Of course, we can only talk about such
an expression after analytic continuation, since the former is defined
for $|z_1|>>|z_2|$, and the latter for $|z_1|<<|z_2|$. However, if we
apply analytic continuation, the answer to the question is positive,
and given by the following theorem.

\proclaim{Theorem 4.1}(see \cite{FR}) Let $x_{i\nu}$ be a basis of
$(V^1)^{\nu-\lambda_0}$, and let $y_{i\nu}$ be a basis of
$(V^2)^{\lambda_2-\nu}$. Then
$$
\gather
R^{V^2V^1}(\frac{z_2}{z_1})^{-1}\Phi^{x_{r\lambda_1},\lambda_1,\lambda_0}(z_1)
\Phi^{y_{s\lambda_1},\lambda_2,\lambda_1}(z_2)=\\
A\sum_{\nu,i,j} E_{ijrs\lambda_1\nu}^{\lambda_2,\lambda_0}
(\frac{z_1}{z_2})^{V^1V^2}
P \Phi^{y_{i\nu},\nu,\lambda_0}(z_2)
\Phi^{x_{j\nu},\lambda_2,\nu}(z_1),\tag 4.1
\endgather
$$
where $A$ is the analytic continuation, $ E^{\lambda,\mu}$ is a
matrix called the matrix of exchange coefficients,
and $P$ is the permutation: $V^1\otimes
V^2\to V^2\otimes V^1$.
\endproclaim

Clearly, the matrix $ E^{\lambda,\mu}(z)$ (we drop the
superscripts $V^1,V^2$ when no confusion is possible)
represents a linear operator
$(V^1\otimes V^2)^{\lambda-\mu}\to(V^2\otimes V^1)^{\lambda-\mu}$. Therefore,
if we define
$$
 E^{\lambda}(z)=\oplus_{\mu} E^{\lambda,\mu}
(z),\tag 4.2
$$
then $ E^{\lambda}(z)$ will correspond to an operator:
 $V^1\otimes V^2\to V^2\otimes V^1$.

Let us introduce some convenient notation. Define the operators
$$
 E_j(z)_{V^1,...,V^n}: V^1\otimes\dots \otimes V^j\otimes
V^{j+1}\otimes\dots\otimes V^n\to V^1\otimes\dots \otimes V^{j+1}\otimes
V^{j}\otimes\dots\otimes V^n
$$
as follows: if $v_i\in V^i$, $1\le i\le n$, and $hv_i=
\chi_i(h)v_i$, $\chi_i\in {\frak h}^*$, $h\in {\frak h}$, then
$$
 E_j(z)_{V^1,...,V^n}(v_1\otimes \dots\otimes v_j\otimes
v_{j+1}\otimes\dots\otimes v_n)=v_1\otimes \dots\otimes
 E^{\lambda_{j+1}}(z)^{V^jV^{j+1}}(v_j\otimes
v_{j+1})\otimes\dots\otimes v_n,\tag 4.3
$$
where $\lambda_j$ are defined by
$$
\lambda_j=(\beta-1)^{-1}\biggl(\sum_{i=1}^n\chi_i\biggr)+\sum_{i=1}^j\chi_i,\
0\le j\le n.\tag 4.4
$$

Let $t_j$ denote the elementary transposition (jj+1) in the symmetric
group $S_n$.

For every $\sigma\in S_n$ set $ E_j(z)^{\sigma}=
E_j(z)_{V^{\sigma(1)},\dots,V^{\sigma(n)}}$.
This operator maps $V^{\sigma(1)}\otimes\dots\otimes V^{\sigma(n)}$
to $V^{t_j\sigma(1)}\otimes\dots\otimes V^{t_j\sigma(n)}$
(we make a convention that for two
permutations $\sigma_1,\sigma_2$
$\sigma_1\sigma_2(j)=\sigma_1(\sigma_2(j))$, $1\le j\le n$, i.e. the
factors in a product of permutations are applied from right to left).

Let us describe some properties of the matrices of exchange
coefficients. First of all, $E_j(z)^{\sigma}$ is not a single-valued
function of $z$, so one should consider the function
$\Cal B_j(\zeta)^{\sigma}=E_j(e^{\zeta})^{\sigma}$.

\proclaim{Proposition} [FR] The functions $\Cal B_j(\zeta)$ have the
following properties:

(i) double periodicity:
$$
\gather
\text{(ia): }\Cal B_j(\zeta+\log p)^{\sigma}=\Cal
B_j(\zeta)^{\sigma};\\
\text{(ib): }\Cal B_j(\zeta+2\pi\sqrt{-1})^{\sigma}=L\Cal
B_j(\zeta)^{\sigma}L^{\prime},\ (\log p=-2t(k+1))\tag 4.5\endgather
$$
where $L,L^{\prime}$ are constant matrices;

(ii) the quantum braid (Yang-Baxter) relation:
$$
 \Cal B_j(\zeta_1-\zeta_2)^{\text{Id}} \Cal B_{j+1}(\zeta_1-\zeta_3)^{t_j}
 \Cal B_j(\zeta_2-\zeta_3)^{t_{j+1}t_j}= \Cal
B_{j+1}(\zeta_2-\zeta_3)^{\text
{Id}}
 \Cal B_j(\zeta_1-\zeta_3)^{t_{j+1}} \Cal B_{j+1}(\zeta_1-\zeta_2)^
{t_jt_{j+1}};
\tag 4.6
$$

(iii) unitarity:
$$
\Cal B_j(\zeta)^{Id}\Cal B_j(-\zeta)^{t_j}=1.\tag 4.7
$$
\endproclaim

{\bf Remarks. }1. Properties (ib), (ii) and (iii)
follow directly from the definition of $B_j$. Property (ia)
follows from the fact that matrix elements of products of
intertwiners satisfy the quantum KZ equations.

2. Statement (i) implies that the matrix elements of $B_j$ must express
in terms of elliptic functions. Frenkel and Reshetikhin
showed that in some special cases these functions are
the elliptic solutions of the quantum Yang-Baxter equation
which occur in statistical mechanics.

 From now on we will be assuming that the matrices of exchange
coefficients are known.

Let us define the fundamental trace $\Cal
F(z_1,...,z_n|s)$.

\proclaim{Definition} The fundamental trace is a function
$\Cal F$ of $z_1,...,z_n,p,s,q$ with values in the space
$\text{End}(V^1\otimes\dots\otimes V^n)$
defined by the property:
if $v=v_1\otimes v_2\otimes\dots\otimes v_n$
is a vector in $V^1\otimes\dots\otimes V^n$, $v_i\in V^i$,
$1\le i\le n$, and $hv_i=
\chi_i(h)v_i$, $\chi_i\in {\frak h}^*$, $h\in {\frak h}$, then
$$
\Cal Fv=\text{Tr}\mid
_{M_{\lambda,k}}(\Phi^{v_1,\lambda_1,\lambda_0}(z_1)\dots
\Phi^{v_n,\lambda_n,\lambda_{n-1}}(z_n)Bs^{-D}),\tag 4.8
$$
where $\lambda_j$ are defined by (4.4).
\endproclaim

It follows from Section 3 that the fundamental trace satisfies $p$-difference
equations (3.24). Below we will show that it also satisfies another
system of $s$-difference equations.

Indeed, let us carry the operator $\Phi(z_i)$ from left to right. We
have to interchange it with $\Phi(z_{i+1}),\dots,\Phi(z_n)$, then
with $B$ and $s^{-D}$ (this will change $\Phi(z_i)$ to
$(1\otimes B)\Phi(s^{-1}z_i)$), then to shift it to the left side of the
product, using the property $\text{Tr}(ab)=\text{Tr}(ba)$, and then
interchange it with $\Phi(z_1),\dots,\Phi(z_{i-1})$.
This procedure allows us to deduce a difference equation
for $\Cal F$. Before we write it down, let us introduce some notation.

Let
$\sigma_{jm}= t_{m-1}\dots t_{j+1}t_j$, $j<m\le n$, $\sigma_{jm}=t_m\dots
t_{j-2}t_{j-1}$, $1\le m<j$.

\proclaim{Theorem 4.2}
The fundamental trace $\Cal F$ satisfies the relations:
$$
\gather
R_{j+1j}^{V^{j+1}V^{j}}(\frac
{z_{j+1}}
{z_{j}})^{-1}\Cal F(z_1,...,z_j,z_{j+1}...,z_n|s)=
F(z_1,...,z_{j+1},z_j,...,z_n|s)E_j(\frac{z_j}{z_{j+1}})^{\text{Id}};\\
R_{jj-1}^{V^{j}V^{j-1}}   (\frac{s^{-1}z_{j}}{z_{j-1}})   \dots
R_{j1}^{V^jV^{1}}   (\frac{s^{-1}z_j}{z_1})   B_j^{-1}
R_{nj}^{V^nV^j}   (\frac{z_n}{z_j})^{-1}    \dots R_{j+1j}^{V^{j+1}V^{j}}(\frac
{z_{j+1}}
{z_{j}})^{-1}\times\\
\Cal F(z_1,...,z_j,...,z_n|s)=
\\
\Cal
F(z_1,...,s^{-1}z_j,...,z_n|s)E_{j-1}(\frac{s^{-1}
z_j}{z_{j-1}})^{\sigma_{jj-1}}
\dots E_{1}
(\frac{s^{-1}z_j}{z_{1}})^{\sigma_{j1}}B_j^{-1}E_{n-1}(\frac{z_j}{z_n})
^{\sigma_{jn-1}}\dots E_j(\frac{z_j}{z_{j+1}})^{\text{Id}},
\tag 4.9
\endgather
$$
where $B_j$ denotes the action of $B$ in $V^j$.
\endproclaim

Now consider the system of difference equations
$$
\gather
G(z_1,...,s^{-1}z_i,...,z_n)=M_j(z_1,...,z_n)G(z_1,...,z_i,...,z_n),\\
M_j(z_1,...,z_n)=
R_{jj-1}^{V^{j}V^{j-1}}   (\frac{s^{-1}z_{j}}{z_{j-1}})   \dots
R_{j1}^{V^jV^{1}}   (\frac{s^{-1}z_j}{z_1})   B_j^{-1}
R_{nj}^{V^nV^j}   (\frac{z_n}{z_j})^{-1}    \dots R_{j+1j}^{V^{j+1}V^{j}}(\frac
{z_{j+1}}
{z_{j}})^{-1}.\tag 4.10
\endgather
$$
This is one of the versions of the quantum KZ equations. Let
$G_0(z_1,...,z_n|s)$ denote the fundamental solution of this system of
equations -- the solution with values in
$\text{End}(V^1\otimes\dots\otimes V^n)$ such that $G_0\sim\prod_{i=1}^N
 z_i^{\frac{\log M_i}{\log s}}$ as
$z_i/z_{i+1}\to\infty$, where $M_i$ is the limit of $M_i(z_1,...,z_n)$ when
$z_i/z_{i+1}\to\infty$. It follows from the theory of difference
equations that this solution exists and is unique.
It can be represented as an infinite ordered product of $R$-matrices.

Introduce the matrix-valued function
$$
\Cal K(z_1,...,z_n|s)=G_0^{-1}(z_1,...,z_n|s)\Cal
F(z_1,...,z_n|s).\tag 4.11
$$
Theorem 4.2 is equivalent to the statement that this function
satisfies the difference equation
$$
\gather
\Cal K(z_1,...,z_j,...,z_n|s)=\\
\Cal K(z_1,...,s^{-1}z_j,...,z_n|s)E_{j-1}(\frac{s^{-1}z_j}{z_{j-1}})^
{\sigma_{jj-1}}
\dots E_{1}
(\frac{s^{-1}z_j}{z_{1}})^{\sigma_{j1}}B_j^{-1}E_{n-1}(\frac{z_j}{z_n})
^{\sigma_{jn-1}}\dots E_j(\frac{z_j}{z_{j+1}})^{\text{Id}}.\tag 4.12
\endgather
$$
On the other hand, formula (3.24) implies that
the function $\Cal K$ satisfies another difference
equation:
$$
\gather
\Cal K(z_1,...,pz_i,...,z_n|s)=\\
G_0^{-1}(z_1,...,pz_i,...,z_n)X_i(z_1,...,z_n|s)G_0(z_1,...,z_n|s)
\Cal K(z_1,...,z_i,...,z_n|s).\tag 4.13
\endgather
$$

Define the operators $D_j$ acting on $V^1\otimes\dots\otimes V^n$ as follows:
$$
D_j(v_1\otimes\dots\otimes v_n)=\frac{<\lambda_j,\lambda_j>-
<\lambda_{j-1},\lambda_{j-1}>}{2(k+1)}(v_1\otimes\dots\otimes
v_n),\tag 4.14
$$
where $\lambda_j$ are defined by (4.4). From the definition of the
fundamental trace it follows that
$\Cal F(z_1,...,z_n|s)=\Cal F_0(z_1,...,z_n|s)z_1^{D_1}\dots
z_n^{D_n}$, where $\Cal F_0\to 1$ as $s\to 0$ and
$z_i/z_{i+1}\to\infty$.

Let
$$
\Theta(z|p)=\prod_{m\ge 0}(1-p^mz)(1-p^{m+1}z^{-1})(1-p^{m+1}).\tag
4.15
$$
Then the function
$$
\varepsilon(z,\alpha|p)=z^{\alpha}\frac{\Theta(zp^{\alpha}|p)}{\Theta(z|p)}
\tag 4.16
$$
is $p$-periodic: $\varepsilon(pz,\alpha |p)=\varepsilon(z,\alpha |p)$.
Therefore, the function
$$
\Cal K_0(z_1,...,z_n|s)=\Cal
K(z_1,...,z_n|s)\prod_i\varepsilon(z,-D_i|p)
\tag 4.17
$$
is a single-valued meromorphic function in the region $z_i\ne 0$
which still satisfies equation (4.13).

Thus, we have proved the following theorem.

\proclaim{Theorem 4.3}
The function $\Cal K_0$ satisfies a pair of matrix difference
equations --
a $p$-difference equation from the left:
$$
\Cal K_0(z_1,...,pz_j,...,z_n|s)=Y_j(z_1,...,z_n|s)\Cal
K_0(z_1,...z_j,...,z_n|s),\tag 4.18
$$
and an $s$-difference equation from the right:
$$
\Cal K_0(z_1,...,sz_j,...,z_n|s)=\Cal
K_0(z_1,...,z_j,...,z_n|s)U_j(z_1,...,z_n|s),\tag 4.19
$$
where
$$
Y_j(z_1,...,z_n|s)=G_0^{-1}(z_1,...,z_n|s)X_j(z_1,...,z_n|s)G_0(z_1,...,z_n|s),
\tag 4.20
$$
$$
\gather
U_j(z_1,...,z_n|s)=\\
\biggl(\prod_i\varepsilon(z,D_i|p)\biggr)
E_{j-1}(\frac{z_j}{z_{j-1}})^{\sigma_{jj-1}}
\dots E_{1}
(\frac{
z_j}{z_{1}})^{\sigma_{j1}}B_j^{-1}E_{n-1}(\frac{sz_j}{z_n})
^{\sigma_{jn-1}}\dots E_j(\frac{sz_j}{z_{j+1}})^{\text{Id}}
\times \\ \biggl(\prod_i\varepsilon(z,-D_i|p)\biggr)
.\tag 4.21
\endgather
$$
\endproclaim

We know that the function $U_j$ is a product of $p$-periodic
functions, so it is $p$-periodic (elliptic):
$U_j(z_1,...,pz_i,...,z_n)=U_j(z_1,...,z_i,...,z_n)$.
Therefore, we have
\proclaim{Corollary} The function $Y_j$ is $s$-periodic for all $j$:
$$
Y_j(z_1,....,sz_i,...,z_n)=Y_j(z_1,....,z_i,...,z_n).\tag 4.22
$$
\endproclaim
\demo{Proof} Since $U_j$ is $p$-periodic, the function $\Cal
K_0(z_1,...,pz_i,...,z_n)$ satisfies (4.19) as long as
$\Cal K_0(z_1,...,z_i,...,z_n)$ does. Thus, the function
$\Cal
K_0(z_1,...,pz_i,...,z_n)\Cal
K_0(z_1,...,z_i,...,z_n)^{-1}$ has to be $s$-periodic.
But this function is exactly $Y_i$, Q.E.D.
\enddemo

{\heading
\bf 5. Difference equations with elliptic coefficients and double
difference systems
\endheading

In this section we survey the theory of systems of difference
equations, in particular those with elliptic coefficients, and
introduce the notion of a double difference system which naturally
arises from this theory.

Let $p$ and $s$
denote two nonzero complex numbers such that $0<|p|,|s|<1$.
If $f(\bold z)=f(z_1,...,z_n)$ is a matrix-valued
function of $n$ complex variables then let
$$
\gather
P_if(\bold z)=f(z_1,...,pz_i,...,z_n),\\
S_if(\bold z)=f(z_1,...,sz_i,...,z_n).\tag 5.1
\endgather
$$

We will say that a meromorphic function $f$ in $\Bbb C^{*n}$ is
$p$-elliptic (respectively, $s$-elliptic) if $P_if=f$ (respectively
$S_if=f$) for all $i$.

Let $a_i(\bold z)$, $1\le i\le n$ be arbitrary meromorphic
functions in $\Bbb C^{*n}$
with values in $N\times N$ matrices. Consider the system of
difference equations
$$
P_if(\bold z)=a_i(\bold z)f(\bold z),\tag 5.2
$$
where $f$ is an $N\times N$-matrix valued function.

\proclaim{Definition} A system of $p$-difference equations is
called {\it consistent} if there exists a meromorphic solution $f$ to
this system whose determinant is not identically zero.
\endproclaim

The following obvious proposition gives necessary conditions for
consistency of equations (5.2).

\proclaim{Proposition 5.1} If system (5.2) has a nonzero meromorphic
solution $f$ such that $\text{det}(f)$ is not identically zero, then
$$
P_ia_j(\bold z)\cdot a_i(\bold z)= P_ja_i(\bold z)\cdot a_j(\bold
z).\tag 5.3
$$
\endproclaim

\proclaim{Definition} If a system of difference equations
satisfies (5.3), it is called {\it holonomic}.
\endproclaim

Thus, any consistent system must be holonomic.

{\bf Remark. } It is obvious that if system (5.2) admits a meromorphic
solution $f$ whose determinant is not identically zero then
all meromorphic solutions of (5.2) form an $N^2$-dimensional
 vector space over the field of $p$-elliptic functions:
all of them have the form $fg$, where $g$ is a matrix-valued
$p$-elliptic function.

In the special case when the coefficients $a_i$ are rational functions
the above necessary conditions are also sufficient:

\proclaim{Proposition 5.2} Any holonomic system of $p$-difference
equations with rational coefficients is consistent.
\endproclaim

\demo{Proof}
We start with proving a technical lemma.

{\bf Lemma. }If $a_i$ are rational for all $i$ then there exist
positive real numbers $r_j$, $1\le j\le n$, such that the
multiannulus $A=\{(z_1,...,z_n)|r_j|p|\le |z_j|\le r_j,1\le j\le
n\}$ does not contain a singular point of $a_i^{\pm 1}$ for any $1\le
i\le n$.

{\it Proof of the Lemma. } Since $a_i$ are all rational, the singular
set for the collection of functions $\{a_i^{\pm 1},1\le i\le n\}$
is an affine algebraic variety in $\Bbb C^n$. By our agreement it is
not the entire $\Bbb C^n$, therefore it is a subset in some algebraic
hypersurface prescribed by the equation
$P(z_1,...,z_n)=0$, where $P$ is a polynomial:
$P=\sum a_{m_1\dots m_n}z_1^{m_1}\dots z_n^{m_n}$.

Let $r_i$ be defined according to the rule: $r_n=r$,
$r_{j-1}=e^{r_j}$. Denote the corresponding annulus $A$
defined in the statement of the lemma by $A_r$.
Let $T$ be the lexicographically highest nonzero term in
the polynomial $P$ (i.e it has the highest possible degree of $z_1$,
and among the terms with that degree of $z_1$ it has the highest
degree of $z_2$, and so on). Then it is easy to see that
$\lim_{r\to\infty}\frac{P}{T}\mid_{A_r}=1$ (i.e. the highest term
dominates all the others). Therefore, if $r$ is sufficiently large,
the polynomial $P$ cannot vanish on $A_r$ (because $T$ does not).
Thus, all the functions $a_i^{\pm 1}$ are regular in $A_r$, Q.E.D.

Now let us prove our proposition. Pick an annulus $A$ satisfying the
condition of the Lemma. Now make this annulus into an $n$-dimensional
complex torus (abelian variety)
as follows: identify two points $(z_1,...,z_i,...,z_n)$
and $(z_1,...,pz_i,...,z_n)$ of $A$ whenever $|z_i|=r_i$. Denote
the resulting quotient space by $E_p^n$ (it is nothing else but the
$n$-th Cartesian power of the elliptic curve $E_p=\Bbb C^*/\{z\sim
pz\}$). System of
difference equations (5.2) can now be interpreted as a gluing condition for
a rank $N$ holomorphic vector bundle over $E_p^n$, and
meromorphic $\Bbb C^N$-valued solutions of (5.2) can be viewed as
meromorphic sections of this bundle. But it is known from elementary
algebraic geometry that any holomorphic vector bundle over a
smooth projective variety has nonzero meromorphic sections,
and the dimension of the space of meromorphic sections over the
field of rational functions is equal to the rank of the bundle.
In our case it implies that (5.2) has $N$ $\Bbb C^N$-valued solutions
which are linearly independent at a generic point. These solutions can
be combined into an $N\times N$-matrix solution which will have a
nonzero determinant at a generic point Q.E.D.
\enddemo

 However, in general a holonomic system of $p$-difference equations
with meromorphic coefficients does not have to be consistent.
A counterexample already exists for systems with elliptic
coefficients.

{\bf Example. }Consider the system of difference equations:
$$
f(pz,w)=f(z,w),\ f(z,pw)=a(z)f(w),\tag 5.4
$$
where $a$ is a non-constant $p$-elliptic function in $\Bbb C^*$. It is clear
that this system satisfies identities (5.3). Still, let us show
that it does not have a nonzero meromorphic solution.

Suppose that $f$ is a nonzero meromorphic solution of (5.4).
Then for all $w$ but a countable set of them $f(z,w)$ is a meromorphic
function in $z$. Also, the order of $f(z,w)$ as a function of $z$
at any fixed point $z_0$
is an essentially constant function of $w$ -- it has the same value
$d(z_0)$ at all points $w$ except for a countable set of them.
However, the second equation in (5.4) shows that if $z_0$ is a pole of
$a$ of order $k$ then
$$
\text{ord}_{z_0}f(z,p^mw)=\text{ord}_{z_0}f(z,w)-km,\tag 5.5
$$
so for sufficiently large $m$ the order of $f(z,p^mw)$ at $z_0$ will
become less than $d(z_0)$ -- contradiction.

{\bf Remark. } Of course, one can find plenty of nonzero
 solutions of (5.4) with essential singularities, e.g.
$$
f(z,w)=\frac{\Theta(w|p)}{\Theta(wa(z)|p)},
$$
where $\Theta$ is defined by (4.15).

In spite of this example, it is very easy to show the existence of
meromorphic solutions for a single difference equation.

\proclaim{Proposition 5.3} Let $a(z)$ be a meromorphic $N\times N$-matrix
function in $\Bbb C^*$ such that $\text{det}(a)$ is not identically
zero. Then the difference equation
$$
f(pz)=a(z)f(z)\tag 5.6
$$
has a meromorphic solution whose determinant is not identically zero.
\endproclaim

\demo{Proof} Let $r$ be a positive real number with the property:
the function $a(z)$ is defined on the circles $|z|=r$ and $|z|=|p|r$,
and its determinant does not vanish anywhere on these circles.
Let $A=\{ z\in\Bbb C^*|r\ge |z|\ge |p|r\}$ be the annulus
squeezed between these circles, and let
$E_p=A/(z\sim pz\text{ if }|z|=r)$ be the elliptic curve obtained
by gluing the boundaries of the annulus $A$ to each other.
Difference equation (5.6) can now be interpreted as a gluing condition for
a rank $N$ holomorphic vector bundle over $E_p$.
 The rest of the argument is as in
the proof of Proposition 5.2.
\enddemo

Now consider system (5.2) in which the coefficients $a_i$ are
$s$-elliptic. Then the system has a new property: if $f(\bold z)$
is a solution then $S_if(\bold z)$ is a solution as well.
If $\text{det}(f)$ is not identically zero, this property implies that
there exist $p$-elliptic matrix-valued functions $b_i(\bold z)$, $1\le
i\le n$, such that
$$
S_if(\bold z)=f(\bold z)b_i(\bold z).\tag 5.7
$$

We see that there is another system of difference equations with elliptic
coefficients (now $s$-elliptic)
satisfied by $f$. Thus, we are naturally lead to
introduce a new notion of a {\it double difference system}.

\proclaim{Definition} A {\it double difference system} is a
system of difference equations of the form
$$
\gather
P_if(\bold z)=a_i(\bold z)f(\bold z),\\
S_if(\bold z)=f(\bold z)b_i(\bold z),\ i=1,...,n\tag 5.8\endgather
$$
where $f$ is an $N\times N$-matrix valued function, and $a_i$, $b_i$
are meromorphic $N\times N$-matrix valued functions whose determinants are not
identically zero.
\endproclaim

Let us now study double difference systems.

Suppose that $f$ is a solution of (5.8) whose determinant is not
identically equal to zero. Then one must have a consistency condition
$$
a_i(\bold z)f(\bold z)P_ib_j(\bold z)=S_ja_i(\bold z)f(\bold
z)b_j(\bold z),\tag 5.9
$$
since both sides of (5.9) are equal to $P_iS_jf(\bold z)$, according
to (5.8).
The simplest way to satisfy this condition automatically is to set
$$S_ja_i=a_i,\ P_ib_j=b_j,\tag 5.10$$ which is the same as to say that
$a_i$
are
$s$-elliptic and $b_j$ are $p$-elliptic. Besides this, we have the
usual consistency conditions
$$
\gather
P_ia_j(\bold z)\cdot a_i(\bold z)= P_ja_i(\bold z)\cdot a_j(\bold
z),\\
S_ib_j(\bold z)\cdot b_i(\bold z)= S_jb_i(\bold z)\cdot b_j(\bold
z).\tag 5.11\endgather
$$

\proclaim{Definition}
A system of the form (5.8) satisfying consistency conditions (5.10)
and (5.11) is called an elliptic double difference system.
\endproclaim

Elliptic double difference systems are exactly those arising from
systems (5.2) with $s$-elliptic coefficients.
Our discussion shows that they are
the most natural examples of double difference systems.

Note that in an elliptic double diffierence system, the coefficients
of the $s$-difference equations play the role of monodromy (or
connection) matrices for the $p$-difference equations, and vice versa.

Of course, (5.10) and (5.11) are only necessary and by no means
sufficient conditions of consistency of system (5.8).
This is demonstrated by the following proposition giving a necessary
and sufficient condition of existence of a nondegenerate solution for
an elliptic double difference system with constant coefficients.

\proclaim{Definition} We say that the numbers $p,s$ are generic
if for $m,k\in\Bbb Z$ $p^m=s^k$ if and only if $m=k=0$. We say that
$p,s$ are strictly generic if they generate a dense subgroup in $\Bbb C^*$.
\endproclaim

\proclaim{Proposition 5.4} Let $p,s$ be strictly generic, and let
$a_i$ and $b_i$ be constant matrices for
all $i$. Then system (5.8) has a
meromorphic solution $f$ with $\text{det}(f)$
not identically equal to $0$ if and only if there exist invertible
$N\times N$ matrices $R,L$ and diagonal matrices $M_i$, $1\le i\le n$, with
integer entries, such that
$$
a_i=Lp^{M_i}L^{-1},\ b_i=R^{-1}s^{M_i}R,\ 1\le i\le n.\tag 5.12
$$
\endproclaim

\demo{Proof}

{\it Sufficiency. } The function
$$
f(\bold z)=L\prod_{i=1}^nz_i^{M_i}R\tag 5.13
$$
is a solution of (5.8) whose determinant is not identically equal to
zero.

{\it Necessity. } Assume that (5.8) has a meromorphic solution $f$ whose
determinant is not identically zero. Then this solution has to be
holomorphic and nondegenerate everywhere. Indeed, suppose that
$\bold z_0$ is a singularity of $f$.
Then (5.8) implies that for all $k_i,m_i\in\Bbb Z$
$\prod_i(P_i^{k_i}S_i^{m_i})\bold z_0$ is a singularity of $f$ as well.
But since $p,s$ are strictly generic, the set of these points is dense
in $\Bbb C^{*n}$ -- a contradiction which implies the holomorphicity
of $f$. Applying the same argument to $f^{-1}$, we get the nondegeneracy.

Now we need to use a simple lemma from the classical theory of
difference equations:

{\bf Lemma. } The difference equation $f(pz)=af(z)$ has a holomorphic
nondegenerate solution if and only if the matrix $a$ is
diagonalizable, and its entries are integer powers of $p$.
This solution, if it exists,
has the form $Lz^MR$, where $M$ is a diagonal
matrix with integer entries, $L,R$ are invertible matrices,
and $a=Lp^ML^{-1}$.

This lemma together with the commutativity condition $[a_i,a_j]=0$
which follows from (5.11), implies that the matrices $a_i$
simultaneously diagonalize in a certain basis, and their eigenvalues
are integer powers of $p$. In other words, there exists an invertible
matrix $L$ and diagonal matrices with integer entries $M_1,...,M_n$
such that $a_i=Lp^{M_i}L^{-1}$, and
$f=Lz_1^{M_1}\dots z_n^{M_n}R$, from which we get (5.12).
\enddemo

\proclaim{Proposition 5.5} If $p,s$ are strictly generic
then the dimension of the space of solutions of any double difference
system (5.8) (over $\Bbb C$) is less than or equal to $N^2$.
\endproclaim

\demo{Proof} Let $z_0$ be a regular point of the coefficients
of (5.8) and all their $p,s$-translates. Such a point obviously exists
since the singular set has codimension 1. Then $f$ must be regular at
this point, and its value
there determines its value anywhere else, since the $p,s$-translates
of $z_0$ form a dense set in $\Bbb C^{*n}$.
\enddemo

{\bf Remarks. }
1. It is easy to show that Propositions 5.3 and 5.4 are true for generic $p,s$
which are not necessarily strictly generic.

2. The dimension of the space of solutions in Proposition 5.5
can be exactly $N^2$: this happens when $a_i$ and $b_i$ are scalar matrices.

3. It follows from Proposition 5.5 that in the case $N=1$ and generic $p,s$,
if a double difference system has a nonzero solution, it is unique up
to a constant factor.

One of the first major problems in the theory of elliptic double
difference systems is the consistency problem:

\proclaim{Consistency problem I} Classify all sets $\{a_i,b_i,1\le i\le
n\}$ for which (5.8) has a solution whose determinant is not
identically zero.
\endproclaim

Another formulation of this problem is:
\proclaim{Consistency problem II} For a given set of coefficients
$\{a_i,1\le i\le
n\}$ find all possible sets $\{b_i,1\le i\le
n\}$
for which (5.8) has a solution whose determinant is not
identically zero.
\endproclaim

In general, this problem is very difficult, and it is not clear how to
approach it, even in the one variable case.
However, in the case $N=1$ (scalar-valued functions) it can be
solved completely as described below.

The idea is to explicitly present a nonzero solution to the $p$-part
of the elliptic double difference system, and then find the
corresponding functions $b_i$. Then the general form of the
functions $b_i$ is $b_i^*=b_i\frac{S_ih}{h}$, where $h$ is a
$p$-elliptic function.

For simplicity we assume that $n=1$; the case $n>1$ is analogous but
somewhat more technical. Our purpose now is to find a nonzero solution
of the equation $f(pz)=a(z)f(z)$, where $a$ is an $s$-elliptic function.

It is known that any scalar-valued elliptic function can be written as
a constant times a ratio of products of theta functions. More
precisely, for every $s$-elliptic function $a(z)$ of one variable there
exist a unique constant $C\in\Bbb C$, $k\in \Bbb Z$, and a unique finite
set $A\subset\{z\in \Bbb C^*||s|\le |z|<1\}$ with a function $\nu:
A\to \Bbb
Z\backslash\{0\}$,
such that
$$
a(z)=Cz^k\prod_{x\in A}\Theta(z/x|s)^{\nu(x)};\tag 5.14
$$
and conversely, the function $a$ given by (5.14) is $s$-elliptic iff
$\sum_{x\in A}\nu(x)=0$, and $\prod_{x\in A}x^{\nu(x)}=s^{-k}$.

Therefore, it is enough to be able to find meromorphic solutions
to the equations
$$
f(pz)=Cf(z),\ f(pz)=z^kf(z),\ f(pz)=\Theta(z/x|s)f(z) \tag 5.15
$$
-- then a solution of $f(pz)=a(z)f(z)$ can be obtained as a product of
such solutions.

A meromorphic solution of $f(pz)=Cf(z)$ is given by
$$
f(z)=\frac{\Theta(z|p)}{\Theta(Cz|p)},\tag 5.16
$$
and
$$
f(sz)/f(z)=\frac{\Theta(sz|p)\Theta(Cz|p)}{\Theta(Csz|p)\Theta(z|p)}.\tag
5.17
$$
A meromorphic solution of $f(pz)=z^kf(z)$ is given by
$$
f(z)=(-\Theta(z|p))^{-k},\tag 5.18
$$
and
$$
f(sz)/f(z)=\biggl(\frac{\Theta(z|p)}{\Theta(sz|p)}\biggr)^k.\tag 5.19
$$
Finally, a meromorphic solution of $f(pz)=\Theta(z/x|s)f(z)$ is given by
$$
f(z)=\prod_{i,j=0}^{\infty}(1-p^iq^jz/x)^{-1}\prod_{i,j=1}^{\infty}
(1-p^iq^jx/z)\Theta(z|p)\Theta(\phi(s)z|p)^{-1},\tag 5.20
$$
where $\phi(s)=\prod_{m\ge 1}(1-s^m)$ is the Euler product, and
$$
f(sz)/f(z)=\frac{\Theta(z/x|p)}{\phi(p)}\frac{\Theta(sz|p)\Theta(\phi(s)z|p)}
{\Theta(z|p)\Theta(s\phi(s)z|p)}.\tag 5.21
$$

The above formulas allow us to construct a solution of
$f(pz)=a(z)f(z)$ for any elliptic function $a$ as a ratio of products
of expressions  (5.16),(5.18),(5.20), which helps us solve the
consistency problem.

Let $K(p)$ be the field of $p$-elliptic functions, and let $K(p)^*$ be
its multiplicative group. Also, let $K(p)^{\prime}$ be the multiplicative
group generated by constants and the functions $z$, $\Theta(z/x|p)$.
Let $L(p,s)^{\prime}$ be the subgroup of $K(p)^{\prime}$
consisting of functions of the form $f(z)/f(sz)$, $f\in K(p)^{\prime}$.
Let $K(p,s)^{\prime}=K(p)^{\prime}/L(p,s)^{\prime}$, and let
$K(p,s)^*=K(p)^*/(L(p,s)^{\prime}\cap K(p)^*)$.

Let $F: K(s)^{\prime}\to K(p,s)^{\prime}$ be the group
homomorphism defined by:
$$
C\mapsto (5.17),\ z^k\mapsto (5.19), \Theta(z/x|s)\mapsto (5.21),\
|s|\le|x|< 1,\tag 5.22
$$
It is easy to check that $F$ restricts to a homomorphism
$F:K(s)^*\to K(p,s)^*$.
Moreover, it is clear that the kernel of $F$ is $L(s,p)^{\prime}$, so
$F$ gives rise to a map $F_0:K(s,p)^*\to K(p,s)^*$. The map $F_0$ is obviously
a group isomorphism.

Now we can formulate the necessary and sufficient condition for
consistency of system (5.8).
If $a\in K(p)^*$, let $\phi_{ps}(a)$ be the image of $a$ in $K(p,s)^*$.

\proclaim{Proposition 5.6} The system of difference
equations
$$
f(pz)=a(z)f(z),\ f(sz)=f(z)b(z)
\tag 5.23
$$
in which $a$ is $s$-elliptic and $b$ is $p$-elliptic,
is consistent if and only if
$F_0(\phi_{ps}(a))=\phi_{sp}(b)$.
\endproclaim

Finally let us describe an explicit formula for solutions of double
difference systems, which applies in the case when $p,s$ are strictly generic.
Consider the system
$$
f(pz)=a(z)f(z),\ f(sz)=f(z)b(z)
\tag 5.24
$$
where $a,b,f$ are $N\times N$-matrix valued functions,
and $a$ is $s$-elliptic, $b$ is $p$-elliptic. Assume that $f$
is a meromorphic
solution to (5.24), and let $z_0$ be a point at which all matrices
$a(p^mz)$, $b(s^nz)$, $m,n\in \Bbb Z$,
 are regular and nondegenerate. Let $z\in \Bbb C^*$.
Let $\{m_j\}$, $\{n_j\}$ be sequences of integers such that
$\lim_{j\to\infty}p^{m_j}s^{n_j}z_0=z$.
Such sequences can be constructed easily with the help of continuous fraction
expansions.
Then the following formula is valid:

\proclaim{Proposition 5.7}
$$
f(z)=\lim_{j\to\infty}\prod_{j=m_k-1}^{0}a(p^jz_0)f(z_0)\prod_{j=0}^{n_k-1}
b(s^jz_0).\tag 5.25
$$
\endproclaim

Existence of this limit and the fact that it equals $f(z)$ follows
directly from (5.24).

Formula (5.25) determines $f(z)$ up to a finite number of unknown parameters
-- entries of $f(z_0)$. It can be generalized to the case of several variables.

Let us now construct another consistent elliptic double difference
system related to system (4.18), (4.19).

Consider an elliptic double difference system of the form
$$
\gather
P_if(\bold z,p,s)=a_i(\bold z,p,s)f(\bold z,p,s),
S_if(\bold z,p,s)=f(\bold z,p,s)b_i(\bold z,p,s),\tag 5.26
\endgather
$$
where $a_i,b_i,f$ are $N\times N$-matrix valued functions whose determinant is
not identically zero, and let us assume that $b_i
(\bold z,p,ps)=b_i(\bold z,p,s)$.
It is obvious that this condition is satisfied for system (4.18), (4.19).
Assume that $f$ is a solution to the system, and let
$$
c(\bold z,p,s)=f(\bold z,p,ps)f(\bold z,p,s)^{-1}.\tag 5.27
$$
Then we can write two equations for $c$. First of all,
$$
\gather
P_iS_if(\bold z,p,ps)=f(\bold z,p,ps)b_i(\bold z,p,ps)=c(\bold z,p,s)
f(\bold z,p,s)
b_i(\bold z,p,s);\\
P_iS_if(\bold z,p,ps)=P_iS_ic(\bold z,p,s)P_iS_i
f(\bold z,p,s)=P_iS_ic(\bold z,p,s)a_i(\bold z,p,s)f(\bold z,p,s)b_i
(\bold z,p,s),
\tag 5.28
\endgather
$$
which implies:
$$
P_iS_ic(\bold z,p,s)=
c(\bold z,p,s)a_i(\bold z,p,s)^{-1}.\tag 5.29
$$
On the other hand,
$$
P_if(\bold z,p,ps)=a_i(\bold z,p,ps)c(\bold z,p,s)f(\bold z,p,s)=
P_ic(\bold z,p,s)a_i(\bold z,p,s)f(\bold z,p,s),\tag 5.30
$$
from which we have
$$
a_i(\bold z,p,ps)c(\bold z,p,s)=
P_ic(\bold z,p,s)a_i(\bold z,p,s)
\tag 5.31
$$

Equatins (5.29) and (5.31) together imply:
$$
S_ic(\bold z,p,s)=S_ia_i(\bold z,p,ps)c(\bold z,p,s).\tag 5.32
$$

Thus, we have shown that the function $c(\bold z,p,s)$ satisfies a new elliptic
double difference system (5.29), (5.32), which involves the functions $a_i$
and does not involve $b_i$.

Now assume that as $s\to 0$, $f(\bold z,p,s)$ has a finite limit
$f(\bold z,p,0)$
which is known and generically nondegenerate.
(This property holds for system
(4.18),(4.19): the fundamental trace converges to the highest matrix
element of the intertwiner as $s\to 0$).
Then one can write the following formula for $f$:
$$
f(\bold z,p,s)=\prod_{j=0}^{\infty}c(\bold z,p,p^js)^{-1}f(\bold
z,p,0).
\tag 5.33
$$
Thus, if we knew an explicit expression for $c$, we could get a more
 structured expression for $f$ than that coming from (5.25),
which would also be free from unknown parameters.

Unfortunately, for system (4.18), (4.19) it is not clear how to
compute $c$ explicitly; however, the quasiclassical limit of $c$
can be understood. This quasiclassical limit, i.e.
the first term of the Taylor expansion of $c$
near $q=1$ (where $f=\Cal K_0$) is equal to the
right hand side coefficient of the ``moduli equation''
for the fundamental trace of the classical affine Lie algebra
(\cite{E}, Eqn. (3.24)) -- the equation characterizing the derivative
of the fundamental trace with respect to the modular parameter
of the corresponding elliptic curve. This fact is reassuring, since it
gives rise to a hope that the method of \cite{E} can somehow be
generalized to the quantum case, which would allow one to produce
some kind of explicit expression for the function $c$, and hence for
the fundamental trace $\Cal F$.

\heading
{\bf Appendix: limiting cases}
\endheading

Let us briefly describe some
interesting limiting cases. These cases correspond
to some special (limiting) values of the parameters
$p,q,s$.

{\it Case 1.} The quantum KZ limit: $s=0$. In this case, the fundamental
trace transforms into the highest matrix element of a product of
intertwiners, system (3.24) becomes the quantum KZ system of
Frenkel and Reshetikhin, and relations (4.9) become the connection
relations for the quantum KZ system. In this limit, one gets
$q$-hypergeometric functions and their generalizations.

{\it Case 2.} The elliptic KZ limit: $q\to 1$, $p=q^{-2(k+1)}$, $k$ is
fixed. In this case, the fundamental trace becomes
the fundamental trace for the classical affine Lie algebra,
system (3.24) degenerates to the elliptic $r$-matrix system involving
Belavin's elliptic $r$-matrix, and relations (4.9) degenerate to the
monodromy relations for the elliptic $R$-matrix equations (see \cite{E}).
In this limit, one gets transcendental functions of an elliptic curve,
vector-valued modular forms etc.

{\it Case 3.} The Yangian limit: $q=e^{\varepsilon}$, $p=q^{-2(k+1)}$,
$k$ is fixed, $z_i=e^{\varepsilon x_i}$, $\varepsilon\to 0$.
In this case, the degeneration of the fundamental trace
should be something like the fundamental trace for the (doubled)
Yangian of $\frak{sl}_N$, and the equations (3.24)
should converge to a trigonometric deformation of the
Smirnov's equations \cite{cf [LS]}
This limit is still unexplored.

{\it Case 4.} The critical limit: $p\to 1$. In this case,
equations (3.24) transform into an elliptic analogue of
the Bethe ansatz equations. The Bethe ansatz equations are
obtained if one combines this limit with the limit $s\to 0$
(cf. \cite{TV}).

Besides these, there are many other unexplored
limiting cases which are easier to study than the general case.
It is expected that studying these limiting cases,
one should be able to get interesting information
about various classes of special functions arising in representation
theory of Lie algebras and quantum groups.

\end